# A Model Approach to Build Basic Ontology

Debajyoti Mukhopadhyay[1], Sajeeda Shikalgar[2]
Department of Information Technnology[1,2]
Maharashtra Institute of Technology
Pune 411038, India
{debajyoti.mukhopadhyay, Sajeeda.dsr}@gmail.com

**Abstract:** As today's world grows with the technology on the other hand it seems to be small with the World Wide Web. With the use of Internet more and more information can be search from the web. When Users fires a query they want relevancy in obtained results. In general, search engines perform the ranking of web pages in an offline mode, which is after the web pages have been retrieved and stored in the database. But most of the time this method doesn't provide relevant results as most of the search engines were using some ranking algorithms like page Rank, HITS, SALSA and Hilltop. Where these algorithms doesn't always provides the results based on the semantic web. So a concept of Ontology is been introduced in search engines to get more meaningful and relevant results with respect to the user's query.Ontologies are used to capture knowledge about some domain of interest. Ontology describes the concepts in the domain and also the relationships that hold between those concepts. Different ontology languages provide different facilities. The most recent development in standard ontology languages is OWL (Ontology Web Language) from the World Wide Web Consortium (W3C). OWL makes it possible to describe concept to its full extent and enables the search engines to provide accurate results to the user.

**Keywords:** OWL, Protégé, Semantic Web, Query, Classes, Properties, Individuals, XML, Crawler, RDF.

## 1. Introduction

The accelerated processes of digitalization and globally connected databases sprouting that are occurring in recent years have changed the focus of the information problems. It is no longer difficult to find information and gain knowledge about a certain topic, but rather to select from the huge heap of information the most relevant elements only. Search engine traditionally utilize a **syntactic** approach, searching for keywords, and performing operations on their abundance in order to rank the information elements. These methods suffer from problems such as vocabulary inconsistency – a situation in which a certain information object contains relevant information but is not retrieved because it uses different words to describe it – and its "opposite" in which irrelevant information is retrieved due to similarity of words.

Lately however, a new approach is emerging – the **semantic** approach. This approach aims to use meta-data – data about data – in order to answer the users' requirement in a more satisfactory way for Data Retrieval and navigation .
.
Ontologies can be very useful in improving the process in two ways:
1. It allows to abstract the information and represent it explicitly- highlighting the concepts and relations and not the words used to describe them.
2. Ontologies can possess *inference functions*, allowing more intelligent retrieval.
For example a "Tennis player" is also a "professional athlete", and an Ontology that defines the relations between these concepts can retrieve one when the other is queried.

In the field of Information technology Ontology word can be described as intelligent information integration, information retrieval on the Internet, and knowledge management. Ontologies are of basic interest in many different fields, largely due to what they promise: a shared and common understanding of some domain that can be the basis for communication ground across the gaps between people and computers. They (Ontology approaches) allow for sharing and reuse of knowledge bodies in computational form. As many traditional activities are changing their manner in the world of today due to the availability of information brought by the World-Wide-Web (WWW), Ontologies are likely to change more when the knowledge is structured in machine readable way, and the abstracts concepts it contains are shared.

Our proposed model attempts to bring a short and a brief survey of the way experts define Ontology. And here we also going to attempt to design ontology in its one of the best form. Ontologies play a dominant roles in a growing number of different fields. A few examples like in natural language applications (like wordnet) [1], Database and information retrieval areas (like SEMEDA) [2].

## TYPES OF ONTOLOGIES

There are three main types of Ontologies are there to deal with the information on the web. They are summarized in the table below:

| 1. Domain Ontologies | Designed to Represent knowledge relevant to a certain domain type, e.g. medical, mechanical etc. |
|---|---|
| 2. Generic Ontologies | Can be applied to a variety of domain types. Ontologies are applicable to many technical domains. Also Called "super theory" and "core technology". |
| 3. Representational Ontologies | These formulate general representation entities without defining what should be represented. The Frame Ontology is a well known example. |

**Table 1:** Types of Ontologies

---

[1] http://www.wordnet.princeton.edu/
[2] http://www.ncbi.nlm.nih.gov/pubmed/12542408

In our proposed model we are attempting to design and develop Domain Ontologies based on Health care Domain. Where OWL has a richer set of operators - e.g. intersection, union and negation. It is based on a different logical model which makes it possible for concepts to be defined as well as described. Complex concepts can therefore be built up in definitions out of simpler concepts. Furthermore, the logical model allows the use of a reasoner which can check whether or not all of the statements and definitions in the ontology are mutually consistent and can also recognize which concepts fit under which definitions. The reasoner can therefore help to maintain the hierarchy correctly. This is particularly useful when dealing with cases where classes can have more than one parent.

**Types of OWL**

OWL ontologies may be categorized into three species or sub-languages: OWL-Lite, OWL-DL and OWL Full. A defining feature of each sub-language is its expressiveness. OWL-Lite is the least expressive sub-language. OWL-Full is the most expressive sub-language. The expressiveness of OWL-DL falls between that of OWL-Lite and OWL-Full. OWL-DL may be considered as an extension of OWL-Lite and OWL-Full an extension of OWL-DL.

**OWL – Lite:** OWL-Lite is the syntactically simplest sub-language. It is intended to be used in situations where only
a simple class hierarchy and simple constraints are needed. For example, it is envisaged that OWL-Lite will provide a quick migration path for existing theory and other conceptually simple hierarchies.

**OWL –DL:** OWL-DL is much more expressive than OWL-Lite. OWL-DL and OWL-Lite are based on Description Logics (hence the suffix DL). Description Logics are a decidable fragment of First Order Logic and are therefore amenable to automated reasoning. It is therefore possible to automatically compute the classification hierarchy and check for inconsistencies in an ontology that conforms to OWL-DL.

**OWL-Full:** OWL-Full is the most expressive OWL sub-language. It is intended to be used in situations where very high expressiveness is more important than being able to guarantee the decidability or computational completeness of the language. It is therefore not possible to perform automated reasoning on OWL-Full ontologies.

**Components of OWL Ontologies**
   OWL ontology consists of components like Individuals, Properties, and Classes.
**Individuals:** Individuals, represent objects in the domain that we are interested in.

**Properties:** These are binary relations on individuals - i.e. properties link two individuals together. For example hasSibling, hasChild, hasOwner, isOwnedBy are some of the properties that we use while creating OWL ontology.

**Classes:** OWL classes are interpreted as sets that contain individuals. They are described using formal (mathematical)

descriptions that state precisely the requirements for membership of the class. In OWL classes are built up of descriptions that specify the conditions that must be satisfied by an individual for it to be a member of the class.

In this paper, we present an ontology-based framework for extraction and retrieval of semantic information in limited domains. We applied the framework in real time Health care Domain by extracting its web page information and web links and observed the improvements over classical keyword-based approaches by implementing OWL –Lite ontology.

We used Protégé[3], an open source ontology editor and knowledge-based framework that supports two approaches to ontology modeling - (1) Protégé-Frames and (2) Protégé-OWL editors - to design and build the structure of Health care

In our approach for Information Extraction (IE), we identify components in an IE system based on ontologies. Other types of models such as relational models or UML class diagrams can be used for this purpose. However, we believe that ontologies are the best option because of the following reasons.
1. Since ontologies are based on logic, they provide formal mechanisms to define concepts and mappings and support reasoning.
2. The IE systems that follow this approach can be easily converted into OWL files that provide advantages such as the ability to generate semantic contents.

The rest of the paper is organized as follows. Section 2 discusses some related work and section 3 presents the design
of our approach. The details of the results and some discussions we have conducted on this approach are presented in section 4 as Results and Discussions. Sections 5 provides hints of some extension of our approach as future work and conclusion.

## 2. Related Work

The older method of searching relevant results for the query fired by the user were keyword-based information retrieval systems. Where these approaches are based on the vector space model proposed by Salton et al. [1]. In this model, documents and queries are simply represented as a vector of term weights and retrieval is done according to the cosine similarity between these vectors. [2], [3], [4] and [5] are some of the important studies related to traditional searching. This approach does not require any extraction or annotation phase. Therefore, its easy to implement, however, the precision values are relatively low.

The problem of finding and relating cultural heritage information in heterogeneous content with different data format creates an obstacle for end-users and a challenge to research communities. The literature introduces several approaches to ease these problems. (Lynch, 2002) [6] highlights the importance of digitalizing cultural heritage documentation creating Digital Libraries and Digital Collections to make available cultural heritage content. It raises the need for an infrastructure based on a common vocabulary and vocabulary mapping, but out of the Semantic Web.

[3] http://protege.stanford.edu/

(Doerr, 2003) [7] establishes the first ontology for cultural heritage data in collaboration with the International Council of Museums. This high level ontology called CIDOC Conceptual Reference Model is an annotation ontology standard ISO since 2006. It provides an underlying schema composed by over 200 concepts and relations into which other schemas can be transformed, but it does not contain domain ontologies for filling in property values or to detect accessibility issues.

Other approaches like (Benjamins, 2004) [8] extract ontology annotations automatically, integrating different repository contents, but obviating reasoning about them or reflecting accessibility issues.

*Semantic portals* (Hyvönen, 2009) [9] collect contents of various publishers into a single site, based on Semantic Web standards in order to improve structure, extensibility, customization and usability of traditional portal designs. Although they provide reasoning task for recommendations or association discovery, they do not assess accessibility issues since ontology does not model them.

Natural Language Processing (NLP) techniques are important when analyzing text to extract domain ontologies for requirements elicitation. NLP generally refers to a range of theoretically motivated and computational techniques for analyzing and representing naturally occurring texts. The core purpose of NLP techniques is to achieve human-like language processing for a range of tasks or applications [10]. The core NLP models used in this research are part-of-speech (POS) tagging and sentence parsers. POS tagging involves marking up the words in a text as corresponding to a particular part of speech, based on its definition, as well as its context. On the other hand, sentence parsers transform text into a data structure (also called parse tree).Such data structure provides insight into the grammatical structure and implied hierarchy of the input text [11].Instead of using Standford parser/tagger1 and OpenNLP2 we used HTML parser provided by java for this research.

Research on domain engineering is also critical to understand an approach to analyze text with the aim of extracting an ontology for requirements elicitation. Domain engineering highlights the process of reusing domain knowledge in the production of new software systems. Domain engineering particularly aims to support systematic reuse, focusing on modeling common knowledge in a problem domain [12]. Sowa's work on conceptual structures [13] introduces a synthesis of logic, linguistics, and Artificial Intelligence as a mechanism for domain knowledge representation.

RDF Schema for bookmarking and annotating resources [14] are more properly used in The Annotea project[4] from the W3C. The microformats[5] rel-tag [15] and xFolk [16] specifications provide a different approach: Instead of using RDF(S) or OWL, they simply defined additional attributes which can be embedded in any XHTML web page. This relatively simple technique "designed for humans first and machines second" explains why rel-tag and xFolk are already integrated in some folksonomy systems. Though its ease-of-use, it is by far not as powerful as the previously described approaches using Semantic Web technologies.

Our literature survey revealed that current studies on the keyword-based semantic searching are not mature enough:

---

[4] http://www.w3.org/2001/Annotea/
[5] http://www.microformats.org

Either they are not scalable to large knowledgebase's or they cannot capture all the semantics in the queries.

Our aim is to fill this gap by implementing a keyword-based semantic retrieval system using the semantic indexing approach. In other words, we try to implement a system that performs at least as good as traditional approaches and improves the performance and usability of semantic querying. We tested our system in Health care domain to see the effectiveness of semantic searching over traditional approaches and observed a remarkable increase in precision and recall. Moreover we noted that our system can answer complex semantic queries, which is not possible with traditional methods. The study presented in this paper can be extended to other domains as well by modifying the current ontology and the information extraction module as described in [17].

## 3. Proposed Method

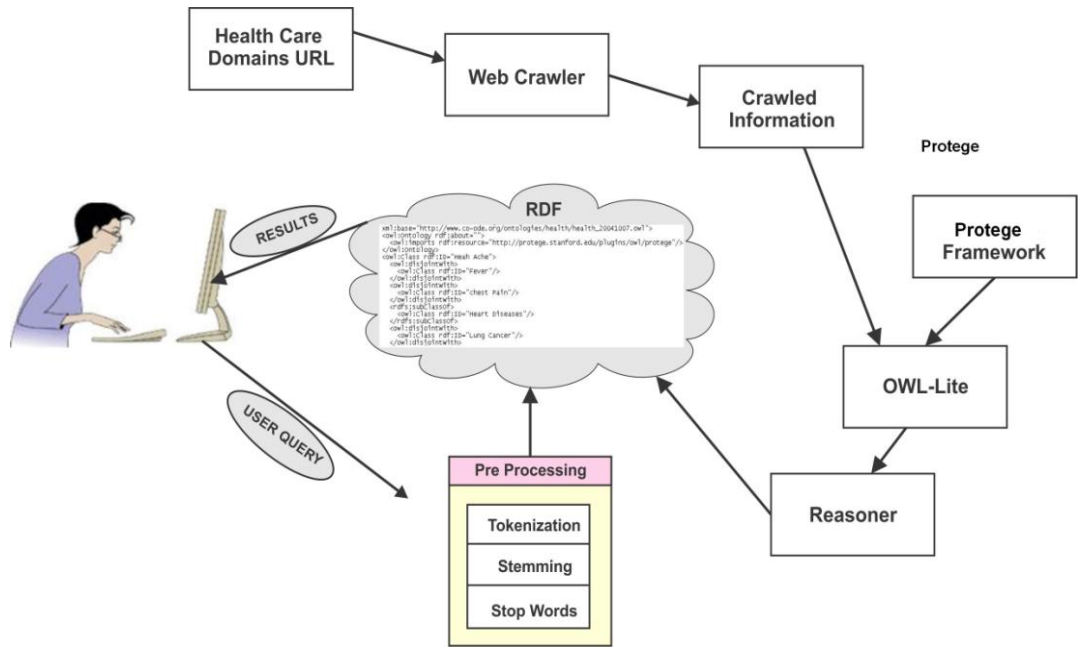

**Fig.1**: Our Proposed Approach

In this section, we describe our approach of creating OWL-Lite Ontology for health care domains according to the steps shown in figure 1. As shown in figure there are 8 main steps in our approach.

**Step 1:** This is the step where we are preprocessing the user query, where query entered by the user is bring down to its basic meaning words by the following four main activities: Sentence Segmentation, Tokenization, Removing Stop Word, and Word Stemming.

Sentence segmentation is boundary detection and separating source text into sentence. Tokenization is separating the input query into individual words. Next, Removing Stop Words, stop words are the words which appear frequently in the query but provide less meaning in identifying the important content of the document such as 'a', 'an', 'the', etc.. The last step for preprocessing is Word Stemming; Word stemming is the process of removing prefixes and suffixes of each word.

**Step 2:** This is the Step where we are collecting health care URLs, Where maximum number of diseases are described properly in the related web pages. We collected the pages from the following web domains like.

http://www.ayushveda.com/
http://www.online-vitamins-guide.com/dietary-cure/dengue-fever.htm
http://www.webmd.com/migraines-eadaches/guide/migraines-headaches-basics

**Step 3:** In this step we are creating a web crawler which accepts a health care URL and searches it's all links.

Web crawlers are an essential component to search engines; running a web crawler is a challenging task. There are tricky performance and reliability issues and even more importantly, there are social issues. Crawling is the most fragile application since it involves interacting with hundreds of thousands of web servers and various name servers, which are all beyond the control of the system.

Web crawling speed is governed not only by the speed of one's own Internet connection, but also by the speed of the sites that are to be crawled. Especially if one is a crawling site from multiple servers, the total crawling time can be significantly reduced, if many downloads are done in parallel.

Despite the numerous applications for Web crawlers, at the core they are all fundamentally the same. Following is the process by which Web crawlers work:

1. Download the Web page.
2. Parse through the downloaded page and retrieve all the links.
3. For each link retrieved, repeat the process.

The Web crawler can be used for crawling through a whole site on the Inter-/Intranet. When we specify a start-URL and the Crawler follows all links found in that HTML page. This usually leads to more links, which will be followed again, and so on. A site can be seen as a tree-structure, the root is the start-URL; all links in that root-HTML-page are direct sons of the root. Subsequent links are then sons of the previous sons [18] [19].

Here in our proposed method we developed a web crawler using java programming language, where we used multithreading feature extensively and also used java html parser to parse the web pages. And finally we store all collected web links in the database.

**Step 4:** In this step we are creating another baby crawler for each and every collected web link of a health care domain. Where this crawler parses the html page and then collects only the web information and stores it in a txt file.

**Step 5:** In this step we are using protégé software to develop OWL-Lite ontology. **Protégé** is a free, open source ontology editor and a knowledge acquisition system. Like Eclipse, Protégé is a framework for which various other projects suggest plugins. This application is written in Java and heavily uses Swing to create the rather complex user interface. Protégé recently has over 160,000 registered users. Protégé is being developed at Stanford University in collaboration with the University of Manchester.

Protégé can be extended by way of a plug-in architecture and a Java-based Application Programming Interface (API) for building knowledge-based tools and applications.

The Protégé platform supports two main ways of modeling ontologies:

- The **Protégé-Frames** editor enables users to build and populate ontologies that are *frame-based*, in accordance with the Open Knowledge Base Connectivity protocol (OKBC). In this model, ontology consists of a set of classes organized in a subsumption hierarchy to represent a domain's salient concepts, a set of slots associated to classes to describe their properties and relationships, and a set of instances of those classes - individual exemplars of the concepts that hold specific values for their properties.
- The **Protégé-OWL** editor enables users to build ontologies for the *Semantic Web*, in particular in the W3C's Web Ontology Language (OWL). "An OWL ontology may include descriptions of classes, properties and their instances. Given such an ontology, the OWL formal semantics specifies how to derive its logical consequences, i.e. facts not literally present in the ontology, but entailed by the semantics. These entailments may be based on a single document or multiple distributed documents that have been combined using defined OWL mechanisms".

**Step 6:** In this Step we are creating OWL –Lite Ontology. Where OWL Lite uses only some of the OWL language features and has more limitations on the use of the features than OWL DL or OWL Full. For example, in OWL Lite classes can only be defined in terms of named super classes (super classes cannot be arbitrary expressions), and only certain kinds of class restrictions can be used. Equivalence between classes and subclass relationships between classes are also only allowed between named classes, and not between arbitrary class expressions. Similarly, restrictions in OWL Lite use only named classes. OWL Lite also has a limited notion of cardinality - the only cardinalities allowed to be explicitly stated are 0 or 1.

The following OWL Lite features related to RDF Schema are included in our approach.
- Class
- rdfs:subClassOf
- rdf:property
- rdfs:subPropertyOf
- rdfs:domain
- rdfs:range
- Individual

**Step 7:** In this step we built classification Hierarchy using reasoner in protégé. Prot´eg´e allows different OWL reasoners to be plugged in, the reasoner shipped with Prot´eg´e is called Fact++. The ontology can be 'sent to the reasoner' to automatically compute the classification hierarchy, and also to check the logical consistency of the ontology. In Prot´eg´e 4 the 'manually constructed' class hierarchy is called the asserted hierarchy. The class hierarchy that is automatically computed by the reasoner is called the inferred hierarchy. To automatically classify the ontology (and check for inconsistencies) the 'Classify...' action should be used. This can be invoked via the 'Classify...' button in the Reasoner drop down menu shown in Figure 2. When the inferred hierarchy has been computed.

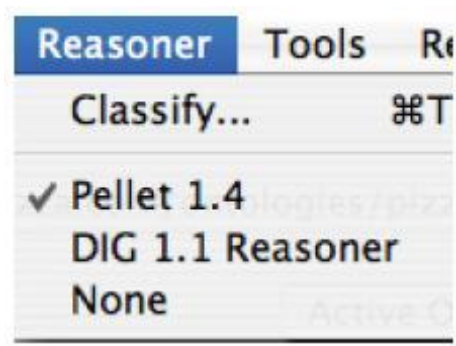

**Fig. 2:** Reasoner menu of protégé 4

**Step 8:** The extensible Markup Language (XML) constitutes the syntactical foundation of the Semantic Web. Nevertheless, this mark-up language is not sufficient to express semantics of Web contents. For this reason, ontologies have been used to express semantics and several languages based on XML (like DAML+OIL, DAML-ONT, OWL and so on) have been proposed to represent ontologies in the framework of the Semantic Web, with increasing expressive power (starting from RDF/RDFS [20]). Ontologies allow to enrich Web contents with semantic annotations to perform reasoning tasks and to obtain new information starting from existing one. Several logical formalisms have been also proposed to make inference on ontologies, searching for a good trade-off between expressiveness and efficiency of reasoning.

This is the last step in our approach where RDF files are being created which helps to answer the preprocessed query of the user, which is been prepared in the step 1. RDF can be used to describe the relationships between entities in the world. RDF can be used at a higher level, too, to describe RDF predicates and classes of resources. Ontologies, schemas, and vocabularies, which all mean roughly the same thing, are RDF information about... other RDF information.

RDF ontologies play a vaguely similar role as XML Document Type Definitions and XML Schema. But they are as different as they are the same. DTDs and XML Schema specify what constitutes a valid document. They don't indicate how a document should be interpreted, and they only restrict the set of elements that can be used in any given file. RDF ontologies, which are themselves written in RDF, provide relations between higher-level things, entirely for the purpose of indicating to applications how some information should be interpreted. RDF ontologies also don't restrict at all which predicates are valid where.

The set of the RDF/S schemas collected were classified under the following two dimensions: (a) the *application domain* they refer to and (b) the *semantic depth* in which they have been developed. **Table 2** presents the collected schemas classified according to these dimensions.

| APPLICATION DOMAIN | SEMANTIC DEPTH |
|---|---|
| Cultural Heritage/Archives/Libraries | Dictionaries and Vocabularies |
| Educational/Academic | Taxonomies |
| Publishing/News | Thesauri |
| Audio-Visual | |
| Geospatial/Environmental: | |
| Biology/Medicine | |
| E-Commerce | |
| Ubiquitous/Mobile/Grid Computing | |
| Cross-Domain | |

**Table 2.** Classification of RDF Schemas according to application domain and semantic depth.

## 4. Results and Discussion

In our proposed method first we collected health care domain URL's using our web crawler and by the same time by using another baby crawler we parse and collect the web page information in txt file format and saves on the disc. By using these Web URL's and collected information with the protégé framework we built OWL and RDF files so that we can easily answer the user's query. The entire health care domain URL's are clustered and saved in the database to bootstrap the search engine even in offline mode. This actually robust our proposed model to provide very accurate results to the users query. An overview of our model can be seen in figure 1.And the results of the proposed model can be seen in figure 3.Where user fires a query to find out the medicine for the **headache**. Then this query is returned with **3** resulted links which are having accurate remedies for the headache.Which were stored in MySQL database and are fetched using ontology in offline mode just as fetching results from Google's Bigtable.

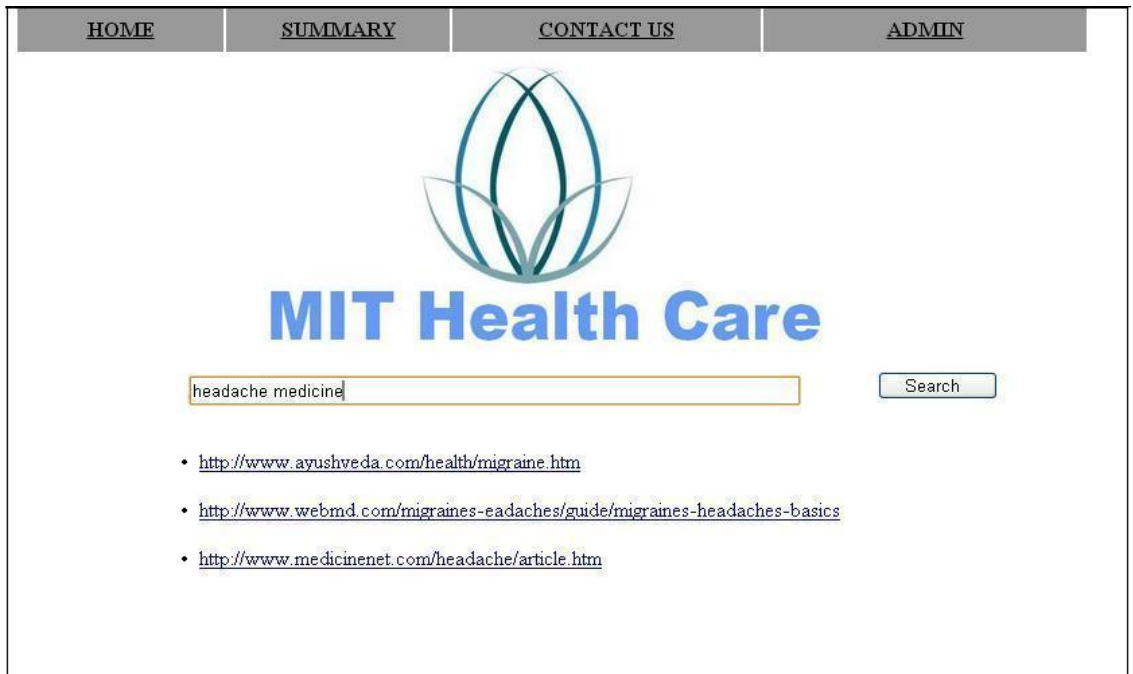

**Fig. 3:** Health care Semantic Search Engine

## 5. Conclusion and Future Work

In this paper, we described an integrated and working Health care search engine system for retrieving semantically enriched remedies for the diseases. The paper presented the implementation of an augmented ontology-based information retrieval system with external open-source resources that uses Protégé to design and build the structure of OWL-Lite ontology as means of retrieving desired URL's for the users query.

  Also, this paper dealt with clustering the augmented links of health care domain. In addition, it explained the mechanism of extracting the information from the web pages and modifying the ontology accordingly by adding the cluster's terms as semantic terms under the "subclass" to which these URL Links belong.

 In our approach we designed a Web crawler using java programming language which helps the model to fetch as many as web links for the given health care domain URL.These links are helping us to build the class hierarchy in ontology so that our system can efficiently provide best result to the users query.

At this point the OWL –Lite ontology provides all the necessary links for respective user query by the entire system. If future designs like OWL-DL and Owl FULL are implemented then this could robust the search engine working speed with catalyzed accuracy.

# References

[1] G. Salton, A. Wong, and C. S. Yang, "A vector space model for automatic indexing," *Commun. ACM*, vol. 18, no. 11, pp. 613–620, 1975.

[2] K. S. Jones, "A statistical interpretation of term specificity and its application in retrieval," *Journal of Documentation*, vol. 28, pp. 11–21, 1972.

[3] G. Salton and M. J. McGill, *Introduction to Modern Information Retrieval*. McGraw-Hill, 1983.

[4] G. Salton, E. A. Fox, and H. Wu, "Extended boolean information retrieval," *Commun. ACM*, vol. 26, no. 11, pp. 1022–1036, 1983.

[5] G. Salton and C. Buckley, "Term-weighting approaches in automatic text retrieval," in *Information Processing and Management*, 1988, pp. 513–523.

[6] Lynch, C., 2002. Digital collections, digital libraries and the digitization of cultural heritage information: *First Monday* 7(6)

[7] Doerr, M., 2003. The CIDOC conceptual reference module: an ontological approach to semantic interoperability of metadata :*AI Magazine*, 24(3)

[8] Benjamins, VR, Contreras, J. et Al., 2004. Cultural heritage and the semantic web: *The Semantic Web: Research and Applications*, p.433-444, Springer.

[9] Hyvönen, E., 2009. Semantic portals for cultural heritage: *Handbook of Ontologies* (2), Springer.

[10] Liddy, E.D., *Natural Language Processing*. 2 ed. Encyclopedia of Library and Information Science. NY. Marcel Decker, Inc., 2001.

[11] Choi, F.Y.Y., *Advances in domain independent linear text segmentation*, in *Proceedings of the 1st North American chapter of the Association for Computational Linguistics conference*. Morgan Kaufmann Publishers Inc.: Seattle, 2000.

[12] Falbo, R.d.A., G. Guizzardi, and K.C. Duarte, *An ontological approach to domainengineering*, in *Proceedings of the 14th international conference on Software engineering and knowledge engineering*. 2002.

[13] Sowa, J.F., *Conceptual structures: information processing in mind and machine*. Addison- Wesley Longman Publishing Co., Inc., 1994

[14] M.-R. Koivunen, R. Swick, J. Kahan, and E. Prud'hommeaux, "An Annotea Bookmark Schema." http://www.w3.org/2003/07/Annotea/ BookmarkSchema-20030707 (viewed 25/06/06), 2003.

[15] T. Celik, "rel-tag, Draft Specification 2005-01-10." http://microformats.org/wiki/rel-tag (viewed 25/06/06), 2005.

[16] B. Gibson, "xFolk (RC1), Draft Specification." http://microformats.org/wiki/xfolk (viewed 25/06/06), 2005.

[17] D. Tunaoglu, O. Alan, O. Sabuncu, S. Akpinar, N. K. Cicekli, and F. N. Alpaslan, "Event extraction from turkish football web-casting texts using hand-crafted templates," in *In Proc. of Third IEEE Inter. Conf. on Semantic Computing (ICSC) (in press)*, 2009.

[18] Franklin, Curt. How Internet Search Engines Work, 2002. www.howstuffworks.com

[19] Grossan, B. "Search Engines: What they are, how they work, and practical suggestions for getting the most out of them," February1997. http://www.webreference.com

[20] G. Klyne and J.J. Carroll, *Resource Description Framework (RDF): Concepts and abstract syntax*, W3C Working Draft, 2003 (http://www.w3.org/TR/2003/WD-rdf-concepts-20030123).

*****